\documentclass[sigconf]{acmart}

\usepackage{multirow}
\usepackage{makecell}
\usepackage{stfloats}
\usepackage{graphicx}
\usepackage{balance}

\copyrightyear{2026}
\acmYear{2026}
\setcopyright{cc}
\setcctype{by}
\acmConference[CIKM '26]{Proceedings of the 35th ACM International Conference on Information and Knowledge Management}{November 07--11, 2026}{Rome, Italy}
\acmBooktitle{Proceedings of the 35th ACM International Conference on Information and Knowledge Management (CIKM '26), November 07--11, 2026, Rome, Italy}
\acmDOI{10.1145/3799682.3841135}
\acmISBN{979-8-4007-2539-5/2026/11}


\author{Abdulaziz Samra}

\affiliation{%
\institution{Skolkovo Institute of Science and Technology}
\city{Moscow}
\country{Russian Federation}
}
\affiliation{%
\institution{Higher School of Economics}
\city{Moscow}
\country{Russian Federation}
}
\email{abdulaziz.samra@skoltech.ru}

\author{Nikita Sukhorukov}

\affiliation{%
\institution{AXXX}
\city{Moscow}
\country{Russian Federation}
}
\affiliation{%
\institution{Skolkovo Institute of Science and Technology}
\city{Moscow}
\country{Russian Federation}
}
\affiliation{%
\institution{Higher School of Economics}
\city{Moscow}
\country{Russian Federation}
}

\email{niksukhorukov7@gmail.com}

\author{Alexey Grishanov}

\affiliation{%
\institution{Sber AI Lab}
\city{Moscow}
\country{Russian Federation}
}

\email{grishanov.av@phystech.edu}

\author{Alexey Vasilev}

\affiliation{%
\institution{Sber AI Lab}
\city{Moscow}
\country{Russian Federation}
}
\affiliation{%
\institution{Higher School of Economics}
\city{Moscow}
\country{Russian Federation}
}

\email{alexxl.vasilev@yandex.ru}

\author{Evgeny Frolov}

\affiliation{%
\institution{AXXX}
\city{Moscow}
\country{Russian Federation}
}

\affiliation{%
\institution{Higher School of Economics}
\city{Moscow}
\country{Russian Federation}
}

\email{evfro@live.ru}

\begin{document}

\title{Cross-Domain Sequential Recommendation with Representation-Level User Transfer}


\begin{abstract}
    Cross-domain sequential recommendation aims to improve next-item prediction in a target domain by leveraging user behavior from auxiliary source domains. However, real-world deployment faces two critical challenges: business constraints that isolate interaction data across domains, and user heterogeneity, where many users lack cross-domain activity. This makes existing cross-domain methods difficult to apply. In this paper, we propose a simple and practical framework for cross-domain sequential recommendations that transfers information exclusively through learned user representations, without sharing raw interaction data or item embeddings across domains. We introduce a lightweight adapter that projects the source-domain user representation into the target-domain embedding space and integrates it with the target interaction sequence. Importantly, when source data is unavailable (i.e., the user observed only in the target domain), the model seamlessly defaults to a robust target-only sequential recommender, retaining benefits from joint training across all users. Extensive experiments were conducted under two setups; one targets benchmarking against previous baselines, while the other focuses on realistic and scalable conditions. Both sets of experiments demonstrate consistent performance gains for shared and unshared users.
\end{abstract}




\maketitle 

\hyphenpenalty=2000        
\exhyphenpenalty=800      
\tolerance=400            
\hbadness=2000            

\section{Introduction}

Sequential recommender systems have become a central component of modern information platforms, enabling personalized prediction of a user’s next interaction based on their historical behavior. Recent advances in deep sequential models, such as Transformer-based architectures, have significantly improved recommendation quality within a single domain~\cite{kang2018selfattentivesequentialrecommendation,sun2019bert4rec,zhou2020s3rec,xie2021cl4srec}. However, in many real-world scenarios, user interaction histories in an individual domain are sparse or incomplete, limiting the effectiveness of even the most expressive sequential models.

Cross-domain sequential recommendation (CDSR) addresses this limitation by transferring knowledge across domains in which the same users interact with different item spaces, such as e-commerce, media consumption, or online services~\cite{chen2024survey_cdsr}. By leveraging behavioral signals from a source domain, CDSR methods seek to improve recommendation performance in a target domain, particularly for users with limited interaction histories. Existing CDSR approaches commonly rely on shared item spaces, joint training on merged interaction data, or pretraining and fine-tuning strategies. While effective in controlled settings, these methods often require strong assumptions that are difficult to satisfy in practice, such as user/item alignment across domains or unrestricted access to raw interaction data~\cite{liu2025llm4cdsr,zhao2023coast,yin2024learningpartiallyaligneditem}. Moreover, parameter-transfer approaches may suffer from negative transfer or catastrophic forgetting when domain distributions differ substantially~\cite{autocdsr}.

Practical deployment of CDSR systems, therefore, faces two fundamental challenges. First, interaction data transfer across domains is often prohibited due to storage or data ownership concerns, making centralized modeling of raw interaction sequences from different domains infeasible~\cite{ppcdr,zhang2025comprehensivesurveycrossdomainrecommendation}. We formalize this setting as a \emph{vertical federated learning} (VFL) problem~\cite{yang2019federated}, where domains possess distinct feature spaces (item sets) but partially overlapping users, and only intermediate representations rather than raw interaction sequences may be exchanged. Second, user overlap between domains is typically partial, and many users may have no interaction history in the source domain at all~\cite{zhu2022ptupcdr}. Effective CDSR models must consequently operate under \emph{restricted data sharing assumptions} and remain robust to \emph{missing source-domain information}.  

To address these challenges, we propose FLADRec (\textbf{F}ederated \textbf{L}earning \textbf{Ad}apted \textbf{Rec}ommender), a representation-level cross-domain sequential recommendation framework. Our key idea is to transfer cross-domain knowledge \emph{exclusively through learned user representations}, rather than raw interactions or shared item embeddings. In the proposed framework, the source-domain encoder produces a fixed-dimensional user representation from its interaction sequence. 

To enable effective transfer, we introduce a lightweight fusing mechanism that conditions target-domain sequence modeling on a projected source-domain user representation. The transferred representation is aligned to the target embedding space and injected directly into the target sequence, allowing the target encoder to \emph{adapt its internal dynamics based on cross-domain user preferences}. This design enforces strict separation between domains while enabling early and flexible integration of source-domain information.

Importantly, the proposed framework naturally supports users without source-domain interaction histories. In such cases, the model reduces to a standard single-domain sequential recommender, while still \emph{benefiting from joint training with users for whom cross-domain information is available}. As a result, the approach improves recommendation quality not only for shared users, but also for users observed \emph{exclusively in the target domain}.

We conduct extensive experiments on multiple cross-domain sequential recommendation benchmarks, demonstrating that the proposed method consistently outperforms strong single-domain and cross-domain baselines.

Our main contributions are:
\begin{itemize}
    \item We propose FLADRec, a framework for CDSR based purely on representation-level transfer. FLADRec transfers only fixed-dimensional user embeddings, without merging raw interaction sequences across domains, enabling deployment under vertical federated learning constraints.
    \item We design an effective conditional fusion mechanism that integrates projected source-domain user representations into target-domain sequence modeling, enabling adaptive preference transfer under strict data separation.
    \item We demonstrate the capability of FLADRec to enhance recommendations for all target users, including those with no source-domain history through implicit regularization, even without explicit cross-domain transfer at inference.
\end{itemize}
\section{Proposed Approach}

\begin{figure*}[htp]
  \centering
  \includegraphics[width=0.95\textwidth]{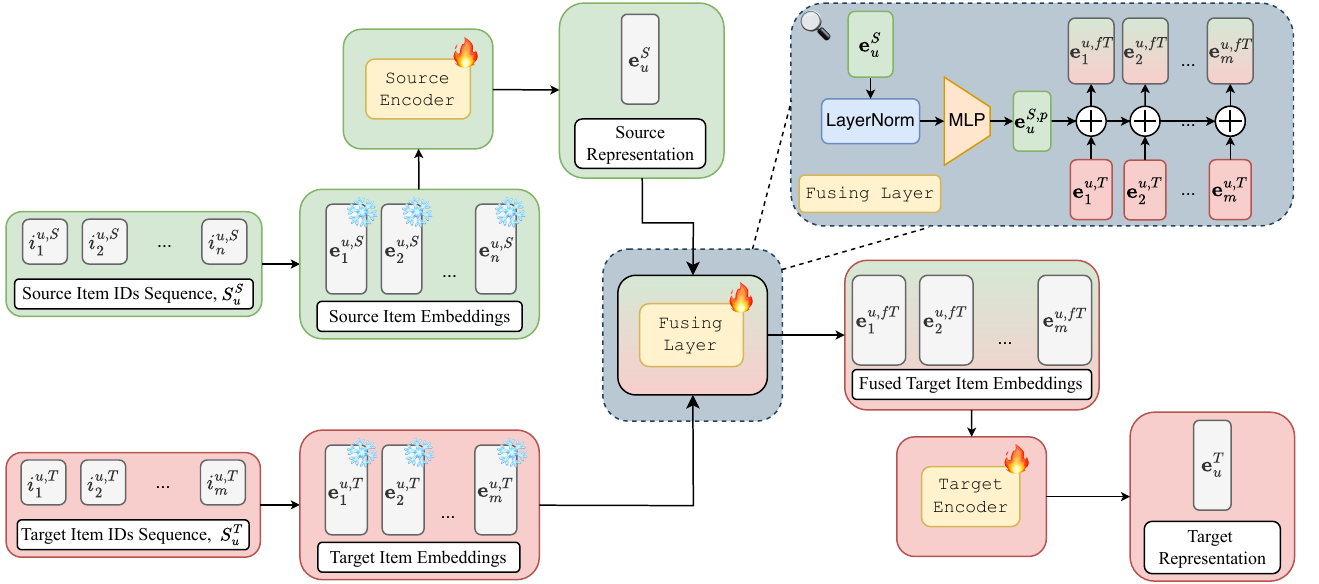}
  \caption{The FLADRec architecture. The Source Domain (Green, upper) and Target Domain (Red, lower) streams process user sequences in parallel. Fusing Layer maps the source sequence representation $\mathbf{e}_{u}^S$ to the target space as $\mathbf{e}_{u}^{S,p}$. This projected vector is then fused element-wise with the target item embeddings \{$\mathbf{e}_{1}^{u, T}, \dots, \mathbf{e}_{m}^{u, T}$\}, acting as a global user context. The fused embeddings \{$\mathbf{e}_{\dots}^{u, fT}$\} are fed into the target encoder for final prediction $\mathbf{e}_u^T$. Icons denote trainable (Flame) vs. frozen (Snowflake) parameters. Note: If a user has no source history, the fusion step is skipped.}
  \label{fig:full_architecture}
\end{figure*}

Figure \ref{fig:full_architecture} illustrates the proposed cross-domain sequential recommendation framework, highlighting the forward pass of user representations through domain-specific encoders and the cross-domain fusion mechanism. 

\subsection{Problem Description}

The core objective is to transfer user preference signals from a source domain to a target domain while respecting practical constraints on data sharing. In particular, we assume that raw interaction sequences \emph{cannot be exchanged across domains} due to distributed storage or even privacy restrictions; instead, only learned user representations are allowed to be transferred.

Formally, given a user $u$ with an interaction sequence in chronological order (from old to new) in the source domain:

\begin{equation*}
    \mathcal{S}_{u}^S = \{i_1^{u, S}, \dots, i_n^{u, S}\},
\end{equation*}
and a sequence in chronological order in the target domain:
\begin{equation*}
    \mathcal{S}_{u}^T = \{i_1^{u, T}, \dots, i_m^{u, T}\},
\end{equation*}
the task is to predict the users' $u$ next target-domain item $i_{m+1}^{u,T}$. Importantly, while $\mathcal{S}_u^S$ \emph{may be absent for users exclusive to the target domain}, our framework seamlessly handles \emph{both shared and non-shared users}.

\subsection{Encoder-Agnostic User Representations}

To ensure the generality of the framework, we do not assume a specific architecture for the sequential encoders. Any model capable of producing a fixed-dimensional representation of a user’s interaction history can be employed, including state-of-the-art sequential recommenders such as SASRec~\cite{kang2018selfattentivesequentialrecommendation}, BERT4Rec~\cite{sun2019bert4rec}, or other Transformer- or RNN-based architectures~\cite{hidasi2016sessionbasedrecommendationsrecurrentneural}.

In the source domain, the item sequence $\mathcal{S}_u^S$ is first mapped to embeddings $\{\mathbf{e}_1^{u, S}, \dots, \mathbf{e}_n^{u, S}\}$ using a (frozen) source item embedding matrix. These embeddings are processed by the source encoder, yielding a compact source-domain user representation $\mathbf{e}^S_u$.

Similarly, in the target domain, the item sequence $\mathcal{S}^T_u$ is mapped to embeddings $\{\mathbf{e}_1^{u, T}, \dots, \mathbf{e}_m^{u, T}\}$, which serve as inputs to the target-side processing pipeline.

\subsection{Fusing layer for cross-domain transfer}

To inject source-domain knowledge into the target-domain sequence modeling, we introduce a Fusing Layer, as shown in Fig. \ref{fig:full_architecture}. This layer operates solely on the transferable source user representation $\mathbf{e}^S_u$, thereby ensuring that user interactions remain isolated and independent across different domains.

Specifically, the source representation is first normalized using Layer Normalization and then projected through a multi-layer perceptron (MLP) to obtain $\mathbf{e}_u^{S,p}$ which lies in the same latent space as the target item embeddings. This projection enables alignment between source-domain user semantics and target-domain item representations.

The projected source user representation is then added element-wise to each target item embedding:
\begin{equation}
    \mathbf{e}^{fT}_k = \mathbf{e}^{u,T}_k + \mathbf{e}_u^{S, p}, \ \ \ k=1,\dots,m,
\end{equation}
resulting in a sequence of fused target item embeddings $\{\mathbf{e}_1^{fT}, \dots, \mathbf{e}^{fT}_m\}$ that are enriched with cross-domain user information.

These fused embeddings are subsequently passed to the target encoder, which produces the final target-domain user representation $e^T_u$ used for next-item prediction.

\subsection{Handling missing source-domain histories}

A key property of the proposed framework is its ability to jointly improve performance for users with and without source-domain interaction histories.

For users lacking source-domain data, the model naturally reduces to a standard single-domain sequential recommender, as the fusion step is skipped and the target encoder operates on unmodified target item embeddings. Crucially, however, these users still benefit indirectly from the training procedure.

During joint training, the target encoder is exposed to both: fused sequences, where target items are conditioned on transferred source preferences, and unfused sequences, corresponding to users without source histories.

As a result, the target encoder learns a representation space that is simultaneously compatible with conditioned and unconditioned inputs. This shared optimization encourages the model to learn more robust and generalizable sequential patterns in the target domain, improving its ability to model user behavior even when no source-domain information is available.

From a representation learning standpoint, the presence of fused sequences acts as a form of implicit regularization, guiding the target encoder toward preference-aware dynamics without imposing additional constraints at inference time. Consequently, the framework not only supports the partial-overlap scenario, but can also improve recommendation quality for users observed exclusively in the target domain.

\subsection{Vertical Federated Learning}

Although users' embeddings transferred from the source are valuable to enhance recommendations in the target domain, they are originally encoded to solve a different task (usually next-item prediction in the source domain). However, these embeddings may not be optimal for the task of cross-domain knowledge transfer. In our proposed solution, we fine-tune the source encoder to generate task-specific embeddings for fusion in the target domain using a vertical federated learning scheme. Vertical Federated Learning (VFL) refers to the federated learning paradigm in which multiple parties share the same set of users but contain different feature spaces~\cite{yang2019federated,Hardy2017PrivateFL}, which are the item catalogs in our case. 

\begin{figure}[htp]
  \centering
  \includegraphics[width=\columnwidth]{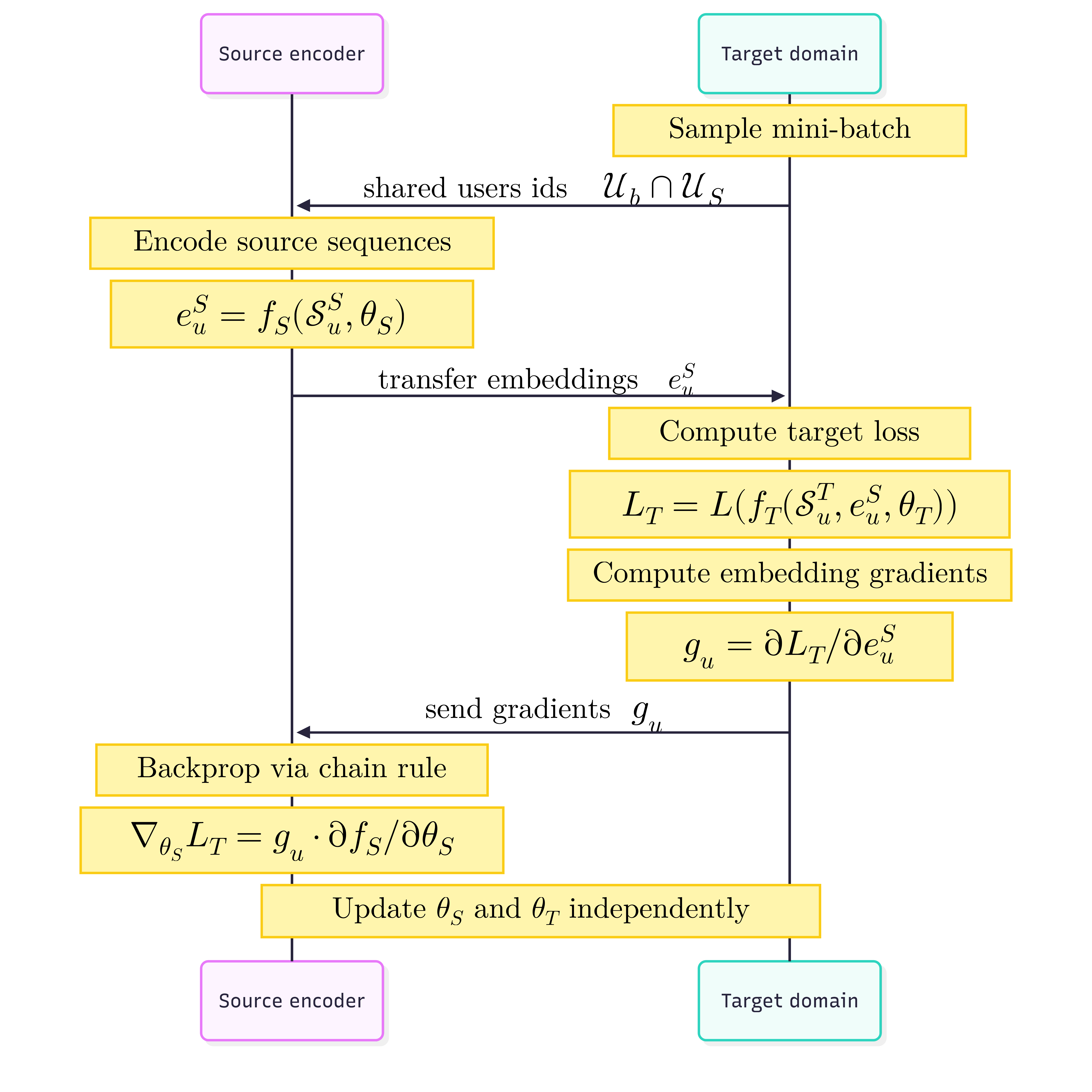}
  \caption{Vertical federated learning protocol for cross-domain sequential recommendation. Throughout, raw interaction data and item IDs remain private to each domain; only embeddings and their gradients are exchanged.}
  \label{fig:federated_learning}
\end{figure}

Let $f_S(\cdot; \theta_S)$ denote the source-domain sequential encoder and $f_T(\cdot; \theta_T)$ the target-domain recommendation model. For a mini-batch of users $\mathcal{U}_b$ sampled from the target dataset, we first encode the source interaction sequences for shared users. The source encoder produces user embeddings
\begin{equation}
    \mathbf{e}_u^{S} = f_S(\mathcal{S}_u^{S}; \theta_S), \quad u \in \mathcal{U}_b\cap \mathcal{U}_S,
\end{equation}
which are transferred to the target domain and used as auxiliary inputs for the target model. The target model optimizes a cross entropy loss \footnote{For large-scale experiments we use approximated version of cross entropy loss~\cite{mezentsev2024scalable}}
\begin{equation}
    \mathcal{L}_T = \mathcal{L}\big(f_T(\mathcal{S}_u^{T}, \mathbf{e}_u^{S}; \theta_T)\big),
\end{equation}
where $\mathcal{S}_u^{T}$ denotes the interaction sequence of user $u$ in the target domain. During backpropagation, gradients with respect to the transferred embeddings are first computed on the target side
\begin{equation}
    \mathbf{g}_u = \frac{\partial \mathcal{L}_T}{\partial \mathbf{e}_u^{S}},
\end{equation}
and then explicitly propagated back to the source encoder by treating $\mathbf{g}_u$ as an external gradient signal. Using the chain rule, gradients of source encoders w.r.t. the target loss are computed as:
\begin{equation}
    \nabla_{\theta_S} \mathcal{L}_T = \mathbf{g}_u \cdot \frac{\partial f_S(\mathcal{S}_u^{S}; \theta_S)}{\partial \theta_S}.
\end{equation}
This gradient-bridging mechanism enables the source encoder to be optimized directly with respect to the target-domain objective, without sharing raw interaction data. Source and target parameters $(\theta_S, \theta_T)$ are updated jointly using independent optimizers, while batches without shared users revert to standard target-only training process. This procedure allows the learned source representations to align with the target recommendation task under a distributed cross-domain training protocol. The overall workflow is illustrated in Figure~\ref{fig:federated_learning}.

\section{Experiment Setup}

In this section we describe the experimental setup. We evaluate our method in two complementary settings to balance \emph{reproducibility against prior work} and \emph{realism in deployment conditions}.

\subsection{Data and Splitting}

We conduct our benchmarking experiments on three datasets used by previous work~\cite{abxi}. These datasets are derived from the Amazon reviews datasets\footnote{\url{https://cseweb.ucsd.edu/~jmcauley/datasets/amazon/links.html}}~\cite{mcauley_2015_image,he_2016_modeling_visual_evolution}, representing three pairs of domains: Food-Kitchen (FK), Beauty-Electronics (BE), and Movie-Book (MB). These datasets serve as established testbeds that enable direct comparison with published baselines under identical preprocessing, user filtering, and evaluation protocols (e.g., sampled metrics with 999 negatives, last-item split, and full user overlap). While essential for reproducibility, such benchmarks \emph{exhibit known limitations}: (i) they are not native cross-domain datasets, interactions originate from a single platform (Amazon), and domain separation is imposed post hoc based on product categories, (ii) artificially high user overlap (only shared users retained), (iii) evaluation via negative sampling—which distorts ranking quality~\cite{krichene_2022_sampled,canamares_2020_sampling}, and (iv) non-temporal or inconsistent data splits that risk temporal leakage~\cite{gusak2025time}. We elaborate further on these pitfalls in Section~\ref{sec:benchmarking_pitfalls}, recommending critical guidelines to advance benchmarking of CDR models.

To address these shortcomings and assess performance under realistic conditions, we additionally evaluate on two industrial-scale, multi-domain datasets: TenRec~\cite{yuan2022TenRec} and Zvuk\&MegaMarket (Zvuk\&MM)~\cite{zvuk-megamarket}. Both capture \emph{authentic cross-domain user behavior} across distinct services (e.g., music streaming and e-commerce) with \emph{naturally partial user overlap} and million-scale item catalogs. Crucially, we employ full-catalog ranking without negative sampling and enforce strict temporal splits, thereby eliminating key biases inherent in benchmarking evaluations. As summarized in Tables~\ref{tab:abxi_data_stats} and~\ref{tab:large_data_stats}, this dual-evaluation strategy allows us to validate competitiveness against prior approaches on \emph{standard benchmarks}, and demonstrate robustness and scalability in settings that \emph{reflect real-world cross-domain recommendation challenges}.


\begin{table}[!t]
\centering

\caption{Statistics of the CDSR benchmark datasets adopted from the same experimental setup from prior work \cite{abxi}.}
\vspace{-0.3cm}
\label{tab:abxi_data_stats}
\small
\resizebox{\columnwidth}{!}{%
\begin{tabular}{c|cccccc}
\toprule
\textbf{Dataset} & \textbf{Users} & \textbf{Items} & \textbf{Inter.} & \textbf{Val Seq.}  & \textbf{Test Seq.} & \textbf{Density} \\
\midrule
Food  & \multirow{2}{*}{7144} & 11837 & 83663 & 2837 & 2419 & 0.098\% \\
Kitchen & & 16258 & 89885 & 4307 & 4725 & 0.077\% \\
\midrule
Beauty & \multirow{2}{*}{4474} & 10379 & 50329 & 2086 & 1875 & 0.108\% \\
Electronics & & 14188 & 63800 & 2388 & 2599 & 0.100\% \\
\midrule
Movie & \multirow{2}{*}{28350} & 35712 & 347650 & 11728 & 10935 & 0.034\% \\
Book & & 90958 & 403147 & 16622 & 17415 & 0.015\% \\
\bottomrule
\end{tabular}
}
\vspace{-0.1cm}
\end{table}
\begin{table}[!t]
\centering
\caption{Statistics of the large-scale CDSR datasets used in the scalability-oriented evaluation. We consider a unidirectional transfer setting, where the first domain in each pair serves as the source and the second as the target.}
\label{tab:large_data_stats}
\resizebox{\columnwidth}{!}{%
\begin{tabular}{lccccc}
\toprule
\textbf{Domains} & \textbf{Users} & \textbf{Items} & \textbf{Inter.} & \textbf{Shared} & \textbf{Overlap} \\
\midrule
QK-Article & 449,288 & 180,951 & 14,887,703 & \multirow{2}{*}{224,644} & \multirow{2}{*}{50.0\%} \\
\hspace{0.5em}$\hookrightarrow$ QK-Video & 449,288 & 694,609 & 11,784,281 & & \\
\cmidrule{1-6}
Zvuk & 95,952 & 822,274 & 38,723,635 & \multirow{2}{*}{47,976} & \multirow{2}{*}{50.0\%} \\
\hspace{0.5em}$\hookrightarrow$ MegaMarket & 95,952 & 453,367 & 4,754,096 & & \\
\bottomrule
\end{tabular}
}
\end{table}


\subsubsection{Benchmarking Datasets}

For Amazon-based benchmarks, we follow ABXI~\cite{abxi} exactly: items with fewer than five interactions are filtered out, user histories are truncated to the most recent 50 interactions, and only users \emph{appearing in both domains are retained}, resulting in 100\% user overlap by construction. The last interaction per user is used for testing, the penultimate for validation, and the remainder for training.

\subsubsection{Scalability-Oriented Datasets}
We filter users with fewer than 10 interactions. To model realistic partial overlap, we randomly subsample non-shared users to achieve a 50\% shared-user ratio across domains, therefore reflecting typical cross-platform scenarios.

\paragraph{Zvuk\&MegaMarket}

For the Zvuk\&MM dataset, we employ a global temporal split strategy. The final week of interaction data is reserved for testing, the penultimate week is utilized for validation, and the remaining historical data comprises the training set. This split ensures there are no data temporal leakage~\cite{gusak2025time}.

\paragraph{TenRec}

As the TenRec dataset lacks explicit timestamps and defines item consumption order solely at the user level, we adopt a user-stratified splitting approach. Each user's history is partitioned sequentially: the initial 80\% of interactions are allocated for training, the subsequent 10\% for validation, and the final 10\% for testing.


\subsection{Training}

\paragraph{Learning Objective.}
We formulate the recommendation task as next-item prediction, where the model predicts a user's next interaction given their historical sequence. For benchmarking datasets, we optimize models using the \emph{full Cross-Entropy (CE)} loss~\cite{klenitsky_2023_dross}. However, for large-scale datasets with extensive item catalogs, full CE becomes computationally infeasible. Therefore, for scalability experiments, we adopt the \emph{Scalable Cross-Entropy} loss~\cite{mezentsev2024scalable} during both the pre-training and adaptation stages.

\paragraph{Initialization and Federated Adaptation.}
The source and target encoder weights are initialized from the \emph{best pretrained single-domain} SASRec models to leverage domain-specific sequential patterns prior to adaptation. To enable cross-domain knowledge transfer without mixing user interactions across domains, we employ vertical federated learning, where the source encoder is fine-tuned using gradients propagated from the target domain.

\paragraph{Optimization and Training Protocol.}
All models are trained using the AdamW optimizer with a batch size of 256. Training proceeds until convergence, with early stopping applied using a patience of 10 epochs based on validation NDCG@10.

The source code and implementation details are publicly available in an anonymous repository\footnote{\url{https://github.com/aa-samra/fladrec}}.

\subsection{Evaluation and Metrics}

We evaluate performance using two standard metrics for sequential recommendation: Normalized Discounted Cumulative Gain (NDCG@$k$) and Hit Ratio (HR@$k$) with $k\in\{10, 50, 100\}$. We also report Coverage@$k$ (COV@$k$) to measure the ratio of the whole item catalog covered by recommended items, providing insight into recommendation diversity and long-tail item promotion.

Following established benchmarking practice for the Amazon datasets~\cite{abxi}, we compute metrics via sampled evaluation with 999 negatives per positive item to enable fair comparison with prior work under identical conditions.

For industrial datasets, to ensure a rigorous assessment, metrics are calculated via full-ranking over the entire item catalog. This rigorous evaluation eliminates sampling bias~\cite{krichene_2022_sampled,canamares_2020_sampling} and provides a more accurate assessment of model capabilities in real-world scenarios where systems must rank millions of items.

To ensure robust and reproducible results, we conduct hyperparameter optimization on the validation split using the Optuna framework with 30 trials per model configuration, optimizing for NDCG@10. Following hyperparameter selection, all models are retrained on the combined training and validation sets to maximize data utilization before final evaluation on the held-out test set.

\subsection{Baselines}

To comprehensively evaluate the proposed framework, we compare its performance against both strong single-domain and representative cross-domain sequential recommendation models. All baselines are implemented under the same data splits, evaluation protocols, and privacy constraints where applicable.

\emph{Single-domain:} we utilize SASRec~\cite{kang2018selfattentivesequentialrecommendation} trained exclusively on the target-domain data as our primary single-domain baseline. We also report BERT4Rec results~\cite{sun2019bert4rec}. For each of these SD models, two settings were reported: (1) Single-Target (ST), where hyperparameters are chosen according to the score computed separately within each individual domain, and (2) Dual-Target (DT), where hyperparameters are selected based on the combined scores obtained across both domains~\cite{abxi}.

\emph{Cross-domain:} FedDCSR~\cite{Zhang_2024_FedDCSR}, which enables federated knowledge transfer between domains through disentangled sequence representation learning. CD-SASRec~\cite{alharbi2021cross} aggregates source-domain sequences into target encoding via dual self-attention. CD-ASR~\cite{alharbi2021d-asr} uses separate self-attention encoders for each domain. MGCL~\cite{xu2025multi} integrates graphical and sequential information under different views and strengthens the profiling via user-to-user contrastive learning on views. C$^2$DSR~\cite{cao2022c2dsr} leverages domain-specific and domain-invariant contrastive tasks. DREAM~\cite{dream} transfers knowledge from domain-specific to cross-domain representations with contrastive learning. ABXI~\cite{abxi} uses task-guided LoRA (Low-Rank Adaptation) modules to extract domain-invariant user interests\footnote{Since no open-source implementations are available for certain baselines (CD-ASR and CD-SASRec), we adopt the evaluation metrics reported in the ABXI study. We ensure that all experiments are conducted on the same datasets and under an equivalent evaluation protocol.}.

\section{Results}

We present a comprehensive evaluation of FLADRec on both standard benchmark datasets and large-scale industrial settings, comparing against strong single-domain and cross-domain baselines under consistent experimental conditions.

\subsection{Benchmarking results}

\begin{table*}[t]
\caption{Benchmark results against models running under limited access to source interaction data}
\vspace{-0.3cm}
\centering
\begin{tabular}{ll|cc|ccc|c}
\toprule
& & \multicolumn{2}{c|}{Single Domain} & \multicolumn{4}{c}{Cross-Domain (Disjoint-data)}\\
\midrule
Domain & Metric & SASRec-1 & BERT4Rec-1 & CD-SASRec & FedDCSR & CD-ASR  & \textbf{FLADRec} \\
\midrule

\multirow{4}{*}{\textbf{Food}}
 & HR@5 & 0.1930 & 0.1819 & 0.1797 & 0.1337  & \underline{0.1976} & \textbf{0.2303} \\
 & HR@10 & 0.2611 & 0.2528 & 0.2454 & 0.2274  & \underline{0.2727} & \textbf{0.3043} \\
 & NDCG@10 & 0.1561 & 0.1462 & 0.1421 & 0.1214  & \underline{0.1616} & \textbf{0.1851} \\
 & MRR & 0.1332 & 0.1230 & 0.1197 & 0.0994  & \underline{0.1368} & \textbf{0.1486} \\
\cmidrule{2-8}
\multirow{4}{*}{\textbf{Kitchen}}
 & HR@5 & 0.1241 & 0.1114 & 0.1119 & 0.0987  & \underline{0.1345} & \textbf{0.2241} \\
 & HR@10 & 0.1851 & 0.1685 & 0.1757 & 0.1454  & \underline{0.1995} & \textbf{0.2891} \\
 & NDCG@10 & 0.1040 & 0.0926 & 0.0946 & 0.0754  & \underline{0.1107} & \textbf{0.1926} \\
 & MRR & 0.0900 & 0.0810 & 0.0821 & 0.0668  & \underline{0.0941} & \textbf{0.1668} \\
\midrule
\multirow{4}{*}{\textbf{Beauty}}
 & HR@5 & 0.1837 & 0.1687 & 0.1605 & 0.1358  & \underline{0.1661} & \textbf{0.2747} \\ 
 & HR@10 & 0.2597 & 0.2438 & 0.2530 & 0.2091  & \underline{0.2550} & \textbf{0.3616} \\
 & NDCG@10 & 0.1523 & 0.1404 & 0.1380 & 0.1061  & \underline{0.1424} & \textbf{0.2135} \\
 & MRR & 0.1295 & 0.1197 & 0.1162 & 0.0872  & \underline{0.1211} & \textbf{0.1696} \\
\cmidrule{2-8}
\multirow{4}{*}{\textbf{Electronics}}
 & HR@5 & 0.1345 & 0.1277 & 0.1290 & 0.1180  & \underline{0.1355} & \textbf{0.2062} \\
 & HR@10 & 0.1894 & 0.1832 & 0.1842 & 0.1708  & \underline{0.1938} & \textbf{0.2627} \\
 & NDCG@10 & 0.1111 & 0.1053 & 0.1069 & 0.0939  & \underline{0.1122} & \textbf{0.1776} \\
 & MRR & 0.0982 & 0.0930 & 0.0948 & 0.0825  & \underline{0.0987} & \textbf{0.1575} \\
\midrule
\multirow{4}{*}{\textbf{Movie}}
 & HR@5 & 0.2258 & 0.2329 & 0.2347 & 0.2154  & \underline{0.2352} & \textbf{0.2678}  \\ 
 & HR@10 & 0.2961 & 0.3105 & 0.3117 & 0.2886  & \underline{0.3052} & \textbf{0.3487} \\
 & NDCG@10 & 0.1647 & 0.1927 & 0.1940 & 0.1866  & \underline{0.1956} & \textbf{0.2260} \\
 & MRR & 0.1874 & 0.1696 & 0.1709 & 0.1743 & \underline{0.1592} & \textbf{0.1977}  \\
\cmidrule{2-8}
\multirow{4}{*}{\textbf{Book}}
 & HR@5 & 0.1357 & 0.1638 & 0.1710 & 0.1265  & \underline{0.1622} & \textbf{0.1859} \\
 & HR@10 & 0.1789 & 0.2152 & \textbf{0.2253} & 0.1578  & 0.2118 & \underline{0.2122} \\
 & NDCG@10 & 0.1007 & 0.1378 & 0.1434 & 0.1030  & \underline{0.1372} & \textbf{0.1716} \\
 & MRR & 0.1147 & 0.1243 & 0.1285 & 0.0929  & \underline{0.1244} & \textbf{0.1661} \\
\bottomrule
\end{tabular}
\label{tab:groupA_transposed}
\end{table*}

\begin{table*}[t]
\caption{Benchmark results against models running under full access to source interaction data}
\vspace{-0.3cm}
\centering
\begin{tabular}{ll|cc|cccc|c}
\toprule
& & \multicolumn{2}{c|}{Single Domain} & \multicolumn{4}{c|}{Cross-Domain (Pooled-data)}& \textbf{Ours}\\
\midrule
Domain & Metric & SASRec-2 & BERT4Rec-2 & MGCL & C$^2$DSR & DREAM & ABXI & \textbf{FLADRec}\\
\midrule

\multirow{4}{*}{\textbf{Food}}
& HR@5 & \underline{0.2313} & 0.2223 & 0.1932 & 0.1984 & 0.2158 & \textbf{0.2498} & 0.2303 \\
& HR@10 & 0.2854 & 0.2956 & 0.2673 & 0.2574 & 0.2771 & \textbf{0.3175} & \underline{0.3043} \\
& NDCG@10 & 0.1797 & 0.1727 & 0.1523 & 0.1546 & 0.1698 & \textbf{0.1973} & \underline{0.1851} \\
& MRR & \underline{0.1535} & 0.1427 & 0.1260 & 0.1311 & 0.1441 & \textbf{0.1679} & 0.1486 \\
\cmidrule{2-9}
\multirow{4}{*}{\textbf{Kitchen}}
& HR@5 & 0.1510 & 0.1363 & 0.1467 & 0.1263 & 0.1377 & \underline{0.1737} & \textbf{0.2241} \\
& HR@10 & 0.2168 & 0.2055 & 0.2157 & 0.1879 & 0.2045 & \underline{0.2410} & \textbf{0.2891} \\
& NDCG@10 & 0.1248 & 0.1116 & 0.1203 & 0.1051 & 0.1138 & \underline{0.1415} & \textbf{0.1926} \\
& MRR & 0.1062 & 0.0948 & 0.1017 & 0.0903 & 0.0956 & \underline{0.1206} & \textbf{0.1668} \\
\midrule
\multirow{4}{*}{\textbf{Beauty}}
& HR@5 & 0.2292 & 0.1970 & 0.1364 & 0.1835 & 0.2090 & \textbf{0.2807} & \underline{0.2747} \\ 
& HR@10 & 0.3262 & 0.3077 & 0.2109 & 0.2645 & 0.3043 & \textbf{0.3835} & \underline{0.3616} \\
& NDCG@10 & 0.1866 & 0.1679 & 0.1162 & 0.1519 & 0.1742 & \textbf{0.2245} & \underline{0.2135} \\
& MRR & 0.1530 & 0.1363 & 0.1001 & 0.1290 & 0.1447 & \textbf{0.1846} & \underline{0.1696} \\
\cmidrule{2-9}
\multirow{4}{*}{\textbf{Electronics}}
& HR@5 & 0.1481 & 0.1519 & 0.1537 & 0.1288 & 0.1216 & \underline{0.1659} & \textbf{0.2062} \\
& HR@10 & 0.2169 & 0.2246 & 0.2159 & 0.1859 & 0.1817 & \underline{0.2389} & \textbf{0.2627} \\
& NDCG@10 & 0.1236 & 0.1275 & 0.1273 & 0.1081 & 0.1023 & \underline{0.1385} & \textbf{0.1776} \\
& MRR & 0.1058 & 0.1101 & 0.1118 & 0.0960 & 0.0895 & \underline{0.1200} & \textbf{0.1575} \\
\midrule
\multirow{4}{*}{\textbf{Movie}}
& HR@5 & 0.2303 & 0.2317 & 0.2097 & 0.2299 & 0.2507 & \textbf{0.2859} & \underline{0.2678} \\ 
& HR@10 & 0.3067 & 0.3095 & 0.2851 & 0.3003 & 0.3255 & \textbf{0.3682} & \underline{0.3487} \\
& NDCG@10 & 0.1903 & 0.1925 & 0.1726 & 0.1911 & 0.2082 & \textbf{0.2388} & \underline{0.2260} \\
& MRR & 0.1673 & 0.1697 & 0.1509 & 0.1700 & 0.1848 & \textbf{0.2118} & \underline{0.1977} \\
\cmidrule{2-9}
\multirow{4}{*}{\textbf{Book}}
& HR@5 & 0.1356 & 0.1547 & 0.1248 & 0.1316 & 0.1469 & \textbf{0.1973} & \underline{0.1859} \\
& HR@10 & 0.1830 & 0.2063 & 0.1668 & 0.1767 & 0.1973 & \textbf{0.2571} & \underline{0.2122} \\
& NDCG@10 & 0.1146 & 0.1302 & 0.1043 & 0.1123 & 0.1237 & \underline{0.1669} & \textbf{0.1716} \\
& MRR & 0.1034 & 0.1176 & 0.0946 & 0.1025 & 0.1118 & \underline{0.1502} & \textbf{0.1661} \\
\bottomrule
\end{tabular}
\label{tab:groupB_transposed}
\end{table*}

In these experiments, we split the compared baselines according to whether each domain has access to the interaction data of the other domain, yielding two groups. The first group comprises single-domain baselines trained independently on each domain’s data, namely SASRec-1 and BERT4Rec-1, as well as the cross-domain models CD-SASRec, FedDCSR, and CD-ASR. In contrast, the second group consists of the same single-domain baselines trained on the union of the two domains’ datasets, denoted as SASRec-2 and BERT4Rec-2. In addition, the other cross-domain models --MGCL, C$^2$DSR, DREAM, and ABXI-- explicitly incorporate mixed raw interaction data from both domains within their pipelines. \emph{We recall that our model belongs to the first category}.

Table~\ref{tab:groupA_transposed} shows the performance comparison of our proposed approach against the first family of baselines on benchmark datasets (Food-Kitchen, Beauty-Electronic, and Movie-Book datasets).\linebreak FLADRec achieves the best overall performance across all dataset pairs and evaluation metrics, achieving the best score in all domains and metrics, with a single exception for the HR@10 on the Book dataset, where it places second. These figures highlight the superiority of FLADRec over other models under the condition of limited access to remote interaction data. 

Table~\ref{tab:groupB_transposed} reports the comparison between FLADRec and the second group of baselines. Although ABXI achieves marginally higher scores on several benchmarks, FLADRec remains broadly competitive despite operating under a more strict regime that prevent interaction‑level data sharing. Notably, the BERT4Rec-2 and SASRec-2 baselines perform slightly worse than the top-performing methods (ABXI and FLADRec), indicating that training strong single‑domain models on mixed datasets is a viable compromise when joint training is permissible; however, this approach forgoes additional gains available via adaptation mechanisms (LoRA in ABXI and source‑encoder fine‑tuning in FLADRec).

\subsection{Scalability Evaluation}\label{sec:scalability}

\begin{table}[htbp]
\centering
\caption{Results of FLADRec model on large-scale datasets. NDCG, HitRate, Coverage (COV) at cutoff $K=10, 50, 100$ and relative improvements w.r.t. the SASRec baseline are reported.}
\label{tab:scalability_results}
\begin{tabular}{c l c c c}
\toprule
\textbf{Dataset} & \textbf{Metric} & \textbf{SASRec} & \makecell{\textbf{FLADRec} \\ \textbf{(Ours)}} & \textbf{Imp.} \\
\midrule
\multirow{10}{*}{\makecell[c]{QK-article \\ QK-video}} 
    & NDCG@10  & 0.0225 & \textbf{0.0283} & +24.1\% \\
    & NDCG@50  & 0.0382 & \textbf{0.0469} & +22.8\% \\
    & NDCG@100 & 0.0469 & \textbf{0.0569} & +21.3\% \\ 
    \cmidrule(l){2-5}
    & HR@10    & 0.0423 & \textbf{0.0532} & +25.8\% \\
    & HR@50    & 0.1155 & \textbf{0.1395} & +20.8\% \\
    & HR@100   & 0.1691 & \textbf{0.2017} & +19.3\% \\
    \cmidrule(l){2-5}
    & COV@10    & 0.1682 & \textbf{0.1794} & +6.6\% \\
    & COV@50    & \textbf{0.2634} & 0.2601 & $-1.2$\% \\
    & COV@100   & \textbf{0.3179} & 0.3009 & $-5.3$\% \\
\midrule
\multirow{10}{*}{\makecell[c]{Zvuk \\ Megamarket}} 
    & NDCG@10  & 0.0583 & \textbf{0.0626} & +7.4\% \\
    & NDCG@50  & 0.0674 & \textbf{0.0739} & +9.6\% \\
    & NDCG@100 & 0.0711 & \textbf{0.0781} & +9.8\% \\
    \cmidrule(l){2-5}
    & HR@10    & 0.0897 & \textbf{0.1063} & +18.5\% \\
    & HR@50    & 0.1306 & \textbf{0.1573} & +20.4\% \\
    & HR@100   & 0.1532 & \textbf{0.1830} & +19.5\% \\
    \cmidrule(l){2-5}
    & COV@10    & 0.1320 & \textbf{0.1320} & 0.0\% \\
    & COV@50    & \textbf{0.2321} & 0.2135 & $-8.7$\% \\
    & COV@100   & \textbf{0.2960} & 0.2600 & $-13.8$\% \\
\bottomrule
\end{tabular}
\end{table}


\begin{table}[ht]
\centering
\caption{Training time (minutes) comparison between ABXI and FLADRec under identical hardware conditions.}
\label{tab:training_time}
\begin{tabular}{c|c|ccc|c}
\toprule
\multirow{2}{*}{\textbf{Dataset}} & \multirow{2}{*}{\textbf{ABXI}} & \multicolumn{3}{c|}{\textbf{FLADRec}} & \multirow{2}{*}{\textbf{Speedup}} \\
\cmidrule{3-5}
& & Pretrain & Adapt & Total & \\
\midrule
Food & \multirow{2}{*}{42.0} & 0.39 & 0.86 & \multirow{2}{*}{3.63} & \multirow{2}{*}{11.57$\times$} \\
Kitchen & & 1.67 & 0.71 & & \\
\midrule
Beauty & \multirow{2}{*}{17.0} & 1.18 & 0.94 & \multirow{2}{*}{3.00} & \multirow{2}{*}{5.67$\times$} \\
Electronics & & 0.33 & 0.55 & & \\
\midrule
Movie & \multirow{2}{*}{65.6} & 5.29 & 2.45 & \multirow{2}{*}{17.82} & \multirow{2}{*}{3.68$\times$} \\
Book & & 5.5 & 4.58 & & \\
\bottomrule
\end{tabular}
\end{table}
We evaluated the computational efficiency of FLADRec on the benchmark datasets used in our main experiments. As shown in Table \ref{tab:training_time}, FLADRec achieves substantial speedups over ABXI, even though it requires 4 stages of training - pre-training two SASRec models and fine-tuning for each transfer.



Given this significant computational boost, which ranges from 3.68$\times$ to 11.57$\times$ across different domain pairs, we further assess FLADRec's scalability by evaluating its performance on two large-scale, real-world cross-domain datasets. Table \ref{tab:scalability_results} reports the results. We compare against SASRec, which serves both as a strong single-domain baseline and as the pretrained initialization for our method.  

Across both TenRec and Zvuk\&MegaMarket, FLADRec consistently outperforms SASRec across all ranking metrics and cutoffs. On TenRec, FLADRec achieves relative improvements in the range (19-24\%) in different metrics and cutoffs. On Zvuk\&MegaMarket, improvements are more moderate in NDCG (7.4–9.8\%) but remain substantial in HitRate, reaching up to 20.4\% at HR@50. 

Importantly, FLADRec also enhances recommendation diversity when generating a small number of candidates: it increases Coverage@10 by +6.6\%, indicating broader exposure to relevant items early in the ranked list. This balance between accuracy and diversity is critical in industrial settings where long-tail item promotion directly impacts user engagement. However, steady declines in coverage are observed at larger cutoffs (@50 and @100), indicating that the recommendations become less diverse when many candidates are generated.

These gains were observed under \emph{partial user overlap}, full-catalog ranking without negative sampling, and strict temporal splits (only Zvuk\&MegaMarket contains timestamps), thereby \emph{avoiding evaluation biases common in benchmark settings}. The consistent improvements indicate that FLADRec enhances recommendation quality in large item spaces with naturally partial user overlap.

Overall, these results support the benchmark evaluation by demonstrating that FLADRec is effectively built on a strong, widely-used single-domain foundation (SASRec) and scales to realistic deployment settings.
\section{Ablation Study}

This section concentrates on the ablation of the components of the proposed approach. In the following ablations, we breakdown the metric computation into three subsets of users (all/shared/unshared). This helps us to understand the effect of each factor in our pipeline on recommendation quality.

\subsection{Effect of cross-domain embedding transfer}

To rigorously evaluate the contribution of source-domain representations to target-domain prediction, we conduct controlled ablation studies where we modify or eliminate the cross-domain signal for shared users. As shown in Figure~\ref{fig:full_architecture}, our framework transfers knowledge exclusively through the representation of the source user $\mathbf{e}^S_u$ at the output of the source encoder. We test three regimes: 
\begin{enumerate}
    \item \emph{Full transfer}, corresponding to the complete FLADRec model; 
    \item \emph{Random injection}, where randomly sampled embedding is injected for all users from the same distribution of $\mathbf{e}^S_u$; 
    \item \emph{No transfer}, where the fusing layer is bypassed entirely during inference, reducing the model to using only target-domain sequences for prediction.
\end{enumerate}

However, unlike training a completely isolated single-domain model from scratch, the target encoder's weights still retain knowledge transferred from the source domain during joint training. This condition isolates the effect of explicit embedding transfer while preserving implicit knowledge transfer through model weights.

    The ablation study results for the tuned source encoder are presented in Table~\ref{tab:ablations}. The full-transfer (full) setting achieves the best performance in both the source-tuned and source-frozen scenarios. When the source encoder is tuned, random embeddings (rand) slightly outperform the no-embedding baseline (none) for the shared subset of users, likely because they introduce user-discriminative signals that complement the adapted encoder weights. Not surprisingly, the metrics in the (full) and (none) settings for the unshared users are identical, since no source embeddings exist to transfer for these users, whereas injecting random source embeddings for unshared users degrades performance. Conversely, when the source encoder is frozen, random embeddings perform worse than no embeddings, as the fixed representation space cannot adapt to the injected noise, leading to uncontrolled perturbations.

\subsection{Ablation of source encoder training}

To investigate the necessity of source encoder adaptation in cross-domain transfer, we conduct a controlled ablation study where we freeze the source encoder weights during training (see Fig.~\ref{fig:full_architecture}). In this configuration, the source representation remains fixed as initially pretrained on its native task, while only the target encoder and fusing layer parameters are updated. This experimental setup isolates the contribution of source encoder fine-tuning to the overall performance gain.

As shown in Table \ref{tab:ablations}, freezing the source encoder leads to consistent performance degradation across all user subsets and transfer variants. Compared to the tuned-source setting, the frozen-source configuration yields lower NDCG@10 for both shared and unshared users, with particularly pronounced drops for shared users under full transfer. These results indicate that projecting fixed, pretrained source representations into the target space—even with an adaptive fusion layer—is insufficient for optimal cross-domain learning.

Overall, this ablation confirms that source encoder adaptation is a critical component of FLADRec, enabling the source representations to align more effectively with the target-domain semantics and thereby enhancing cross-domain transfer. 

Finally, a comparison between the performance improvements obtained by varying the transfer regime (full, random, or none) and those resulting from modifying the source encoder status (trainable versus frozen) indicates that most of performance gain can be attributed to the joint training process of the proposed architecture. 


\begin{table}[t]
\centering
\caption{NDCG@10 comparison across user subsets (ALL / Shared / Unshared). Relative gains w.r.t. baseline are reported.}
\label{tab:ablations}
\resizebox{\columnwidth}{!}{%
\begin{tabular}{l l l c c c}
\toprule
\multicolumn{3}{c}{} & \multicolumn{3}{c}{\textbf{User subset}} \\
\cmidrule(lr){4-6}
\textbf{Model} & \textbf{Regime} & \textbf{Metric} &
\textbf{ALL} & \textbf{Shared} & \textbf{Unshared} \\
\midrule
SASRec & --    & value         & 0.02248 & 0.00976 & 0.01272 \\
\midrule
\multirow{6}{*}{\shortstack{\textbf{FLADRec}\\(Tuned\\Source)}}
  & \multirow{2}{*}{full} & value & \textbf{0.02832} & \textbf{0.01231} & \textbf{0.01601} \\
  &                       & gain,\%  & \textbf{+26.0}   & \textbf{+26.3}   & \textbf{+25.8} \\
  & \multirow{2}{*}{rand} & value & 0.02790 & 0.01226 & 0.01564 \\
  &                       & gain,\%  & +24.5   & +25.7   & +23.0 \\
  & \multirow{2}{*}{none} & value & 0.02814 & 0.01213 & \textbf{0.01601} \\
  &                       & gain,\%  & +25.2   & +24.4   & \textbf{+25.8} \\
\midrule
\multirow{6}{*}{\shortstack{\textbf{FLADRec}\\(Frozen\\Source)}}
  & \multirow{2}{*}{full} & value & 0.02689 & 0.01158 & 0.01531 \\
  &                       & gain,\%  & +19.6   & +18.8   &  +20.3 \\
  & \multirow{2}{*}{rand} & value & 0.02621 & 0.01133 & 0.01488 \\
  &                       & gain,\%  & +16.6   & +16.2   & +17.0 \\
  & \multirow{2}{*}{none} & value & 0.02689 & 0.01158 & 0.01531 \\
  &                       & gain,\%  & +19.6   & +18.8   & +20.3 \\
\bottomrule
\end{tabular}
}
\end{table}

\section{Related Work}

Single-domain recommender systems have achieved significant progress through deep learning architectures like Transformers\cite{kang2018selfattentivesequentialrecommendation,sun2019bert4rec} and self-supervised learning techniques~\cite{zhou2020s3rec,xie2021cl4srec}, which effictively capture the dynamic user preferences. 

For the case of many domains, particularly with partial user overlap, sequential cross-domain recommendation have emerged. These method concentrate on transferring knowledge from source domains to improve recommendation quality in the target domain~\cite{chen2024survey_cdsr} for the next-item prediction task.

A common line of work assumes the presence of users shared across domains and transfers knowledge through shared user representations. Models such as CD-SASRec~\cite{alharbi2021cross}, DCDIR~\cite{bi2020dcdir}, and MIFN~\cite{ma2022mifn} employ domain-aware encoders or attention mechanisms to fuse sequences from multiple domains. More recent CDSR approaches focus on transferring knowledge at the representation level, by aligning latent user embeddings via contrastive objectives or partially sharing model parameters across domains. C$^2$DSR~\cite{cao2022c2dsr} utilized contrastive learning through domain-specific and domain-invariant tasks. ABXI~\cite{abxi} uses task-guided LoRA (Low-Rank Adaptation) modules to identify domain-invariant user interests and finetune the models in both source and target domains. AutoCDSR~\cite{autocdsr} utilizes self-attention to optimize the cross-domain knowledge transfer and combat the negative transfer caused by imbalanced signals from domains. These methods typically rely on joint training over all domains, including the usage of merged interaction sequences across domains and exploit shared users to bridge domain gaps.

The increasing focus on privacy has spurred research into federated recsys, where models are trained across decentralized data sources without exchanging raw user interactions. PPCDR~\cite{ppcdr} proposes a scheme that shares user embeddings protected by local differential privacy technique. However, this method provides learning schemes for domains with full user overlap, which rarely happens in real-world services. FedCDR~\cite{fedcdr} addresses cross-domain recommendation in federated settings but focuses on rating prediction rather than sequential next-item prediction, limiting its applicability to modern sequential recommendation systems.

Our approach bridges critical gaps between cross-domain recommendation and privacy-preserving learning. Unlike prior CDSR methods that require sharing raw sequences or item mappings, our framework transfers knowledge exclusively through distilled user representations, never exchanging raw interactions or embedding tables between domains. Unlike methods that assume full user overlap, our architecture naturally accommodates partial overlap through a fusion mechanism that seamlessly converts to single-domain recommendation when source data is missing. 

Our lightweight adapter enables cross-domain knowledge transfer while maintaining strict domain separation, making it deployable in federated learning scenarios where existing methods fail. Furthermore, our framework uniquely benefits all users, not just those with cross-domain activity. Through joint training with shared parameters, target-domain-only users indirectly benefit from the regularization effect of cross-domain signals, achieving performance gains. This dual benefit represents a practical advance toward deployable cross-domain recommendation systems that respect privacy boundaries while maximizing utility.

\section{Discussion}

\subsection{Privacy Analysis}

FLADRec was essentially designed to achieve cross-domain knowledge transfer between different subsidiaries (services) under a shared ecosystem. The federated training process operates in a distributed storage environment with mild privacy restrictions. Thus, the proposed architecture lack a dedicated mechanism to prevent user history reconstruction attacks. However, we obtain a representation-transfer mechanism that naturally operates effectively whether representations are exchanged via secure channels (VFL) or directly (non-VFL), \emph{without architectural modification}. To secure the training process where gradients are transferred back to the source domain, stronger privacy guarantees, such as formal differential privacy, can be integrated into the federated training protocol.

\subsection{Toward A Fair, Sustainable CDR Benchmarking Protocol}
\label{sec:benchmarking_pitfalls}


    
In this section, we identify and analyze five prevalent pitfalls in benchmarking pipelines for cross‑domain recommendation models, which substantially undermine their practical reliability in industrial settings when compared with single‑domain baselines. As illustrated in our scalability experiments (Section~\ref{sec:scalability}), we took steps to avoid as many of these pitfalls as possible.

\subsubsection{Origin of Data.}
Large companies such as Amazon and Douban release anonymized user-item datasets split into categories, which researchers repurpose to simulate cross-domain recommendation. However, these datasets originate from a single platform, where a shared environment and shared models have access to all data simultaneously. Recently, more "honest" datasets have been released, notably the MegaMarket--Zvuk pair~\cite{zvuk-megamarket} and TenRec~\cite{yuan2022TenRec} are collected from genuinely separate platforms and therefore produced by independent models.

\subsubsection{Domain Overlap Ratio.}
Most CDR research assumes shared users or items across domains and links their representations accordingly. In practice, overlap is limited, creating non-uniform conditions under which models must handle shared and non-shared users differently. Most researchers in the CDR field run experiments on datasets with unrealistically high or even full user overlap, and insufficiently discuss the disparity in quality improvement between shared and unshared users. Notably, recent work has begun addressing quality fairness between shared and non-shared users~\cite{chen2025leave}.

\subsubsection{Dataset Size and Scalability.}
Many recent CDR papers still rely on small, heavily processed datasets ($<$300k interactions), especially when controlling the user overlap ratio. Replicating most such models in real-world settings is typically computationally infeasible. Since recommender systems research is inherently practical, comparisons against baselines evaluated only on small datasets are of limited value, even when those baselines appear to outperform others in that restricted regime.

\subsubsection{Sampled Metric Evaluation.}
Sampled metrics have grown popular because they enable fast computation by ranking relevant items against a random subset of negatives rather than the full catalog. However, this approach can distort model comparisons, and several works have proposed corrections for this bias~\cite{krichene2020sampled}.

\subsubsection{Temporal Leakage in Train/Test Split.}
Sampling each user's most recently interacted item is a common evaluation practice in recommender systems. It is, however, frequently applied to non-temporal or temporally inconsistent data splits, introducing a risk of temporal leakage~\cite{gusak2025time}. Nevertheless, temporal leakage \emph{across} multiple domains remains insufficiently investigated to date.

Specifically, the Zvuk/MegaMarket experiment satisfies all the criteria discussed above, but to a lesser extent with respect to the overlap ratio, as a 50\% user overlap remains relatively high. In addition, the TenRec experiment also does not meet the requirement of a strict temporal split, since the dataset originally lacks absolute timestamps. 

Nonetheless, we argue that enforcing the optimal condition for each of the aforementioned considerations is overly demanding in the context of fast-paced academic research, even though it would promote stronger alignment with industrial applications. We further argue that the partial overlap and scalability conditions are the most critical aspects to prioritize in future CDR research.
\section{Conclusion and future works.}

In this work, we introduced a practical framework for cross-domain sequential recommendation that transfers knowledge through learned user representations, without sharing raw interaction data or item embeddings across domains. Our SASRec-based architecture, combined with a lightweight fusing layer and vertical federated fine-tuning of the source encoder, enables effective alignment of source and target user preferences under realistic privacy and deployment constraints. Extensive experiments across five datasets, including two industrial-scale benchmarks and three established academic benchmarks, demonstrate consistent improvements over strong single-domain and cross-domain baselines, for both shared and target-only users, confirming that representation-level transfer not only avoids negative transfer but also benefits the entire user population. Future work includes extending the framework to multi-source and multi-task settings, exploring alternative fusion mechanisms and backbone architectures, and integrating privacy-preserving mechanisms to protect the exchanged data during learning. 
\section*{Acknowledgments}

This work is an output of a research project (HSE-BR-2025-019) implemented as part of the Basic Research Program at the HSE University.

\section*{GenAI Usage Disclosure}

During the preparation of this manuscript, the authors used Claude to rephrase parts of the text and to improve the readability of the open-source code of this work. All experimental design, data collection, statistical analysis, and interpretation were performed entirely by the authors without the use of GenAI. The authors reviewed and edited all AI-assisted text and code modifications and take full responsibility for the final content of the publication.

\bibliographystyle{ACM-Reference-Format}
\balance
\bibliography{ref}

\end{document}